\documentclass[11pt]{article}

\usepackage[totalheight = 23cm, totalwidth = 17cm]{geometry}
\usepackage{amssymb,amsmath,amsfonts,amsbsy,graphicx,wrapfig}

\def\mathbi#1{\textbf{\em #1}}

\newcommand{\fnl}{f_{\rm NL}}

\newcommand{\calP}{{\cal P}}
\newcommand{\calR}{{\cal R}}

\begin{document}

\begin{titlepage}

\rightline{\footnotesize{CERN-PH-TH/2012-097}} \vspace{-0.2cm}
\rightline{\footnotesize{CPHT-RR 020.0412}}

\begin{center}

\vskip 1.0 cm

{\LARGE  \bf  Heavy fields, reduced speeds of sound and decoupling during inflation}

\vskip 1.0cm

{\large
Ana Ach\'ucarro$^{a,b}$, \hspace{0.2cm} Vicente Atal$^{a}$, \hspace{0.2cm} Sebasti\'an C\'espedes$^{c}$,
\\
Jinn-Ouk Gong$^{d}$, \hspace{0.2cm} Gonzalo A. Palma$^{c}$ \hspace{0.2cm} and \hspace{0.2cm} Subodh P. Patil$^{e}$
}

\vskip 0.5cm

{\it
$^{a}$Instituut-Lorentz for Theoretical Physics, Universiteit Leiden, 2333 CA Leiden, The Netherlands
\\
$^{b}$Department of Theoretical Physics, University of the Basque Country, 48080 Bilbao, Spain
\\
$^{c}$Physics Department, FCFM, Universidad de Chile, Blanco Encalada 2008, Santiago, Chile
\\
$^{d}$Theory Division, CERN, CH-1211 Gen\`{e}ve 23, Switzerland
\\
$^{e}$Centre de Physique Th\'eorique, Ecole Polytechnique and CNRS, Palaiseau cedex 91128, France
}

\vskip 1.2cm

\end{center}

\begin{abstract}

We discuss and clarify the validity of effective single field theories of inflation obtained by integrating out heavy degrees of freedom in the regime where adiabatic perturbations propagate with a suppressed speed of sound. We show by construction that it is indeed possible to have inflationary backgrounds where the speed of sound remains suppressed and slow-roll persists for long enough. In this class of models, heavy fields influence the evolution of adiabatic modes in a manner that is consistent with decoupling of physical low and high energy degrees of freedom. We emphasize the distinction between the effective masses of the isocurvature modes and the eigenfrequencies of the propagating high energy modes. Crucially, we find that the mass gap that defines the high frequency modes increases with the strength of the turn, even as the naive heavy (isocurvature) and light (curvature) modes become more strongly coupled. Adiabaticity is preserved throughout, and the derived effective field theory remains in the weakly coupled regime, satisfying all current observational constraints on the resulting primordial power spectrum. In addition, these models allow for an observably large equilateral non-Gaussianity.

\end{abstract}

\end{titlepage}

\newpage
\setcounter{page}{1}


The recent observation that heavy fields can influence the evolution of adiabatic modes during inflation~\cite{Tolley:2009fg} has far reaching phenomenological implications~\cite{Achucarro:2010jv,Achucarro:2010da,Cespedes:2012hu,Achucarro:2012sm} that, a posteriori, requires a refinement of our understanding of how high and low energy degrees of freedom decouple~\cite{Burgess:2002ub} and how one splits ``heavy'' and ``light'' modes on a time-dependent background. Provided that there is only one flat direction in the inflaton potential, heavy fields (in the present context, field excitations orthogonal to the background trajectory) can be integrated out, resulting in a low energy effective field theory (EFT) for adiabatic modes exhibiting a reduced speed of sound $c_s$, given by 
\begin{equation}\label{speed-of-sound}
c_s^{-2} = 1 + \frac{4 \dot \theta^2 }{  M_{\rm eff}^2 }  \, ,
\end{equation}
where $\dot \theta$ is the  turning rate of the background trajectory in multi-field
space, and $M_{\rm eff}$ is the effective mass of heavy fields,
assumed to be much larger than the expansion rate $H$.  Depending on the
nature of the trajectory, (\ref{speed-of-sound}) can render features
in the power spectrum~\cite{Achucarro:2010da,Cespedes:2012hu} and/or
observably large non-Gaussianity~\cite{Tolley:2009fg,
  Achucarro:2012sm}.

Given that $M_{\rm eff}$ is the mass of the fields we integrate out, one might doubt the validity of the EFT in the regime where the speed of sound is suppressed~\cite{doubt}, as this requires $\dot \theta^2 \gg M_{\rm eff}^2$.
In this article we elaborate on this issue by studying the dynamics of
light and heavy degrees of freedom when $c_s^2 \ll 1$. To this end, we
draw a distinction between isocurvature and curvature field
excitations, and the true heavy and light excitations. We show
that the light (curvature) mode $\mathcal R$ indeed stays coupled to
the heavy (isocurvature) modes when strong turns take place ($\dot
\theta^2 \gg M_{\rm eff}^2$), however, decoupling between the physical
low and high energy degrees of freedom persists in such a way that the
deduced EFT remains valid. 
This is confirmed by a simple setup in which $H$ decreases adiabatically, allowing for a sufficiently long period of inflation. In this construction, {\it high
  energy degrees of freedom} are never excited, and yet {\it heavy
  fields} do play a role in lowering the speed of sound of adiabatic
modes.


We begin by introducing the general setup and notation (see
Refs.~\cite{Achucarro:2010da,Cespedes:2012hu,Achucarro:2012sm} for
details). We consider a non-canonical two-scalar field system with an
action given by 
\begin{equation}\label{total-action}
S=\int \!\! \sqrt{-g}\left[\frac{1}{2} R-\frac{1}{2}g^{\mu\nu} \gamma_{a b}\partial_{\mu}\phi^a\partial_{\nu}\phi^b-V(\phi)\right] \, , 
\end{equation}
(in units $8\pi G=1$) where 
$V$ is the scalar potential  and $\gamma_{a b}$ ($a = 1 , 2$) 
is the sigma model metric of the space spanned by $\phi^{a}$. 
$R$ is the Ricci scalar of an FRW metric $ds^2=-dt^2+a^2(t)\delta_{ij}dx^idx^j$, where $a(t)$ is the scale factor. The background equation of motion for $\phi_0^a(t)$ is then $D_t \dot{\phi}_0^a  +3H\dot{\phi}^a_0+V^a=0$, 
where $H= \dot{a} / a$ and $D_t X^a  = \dot X^a + \Gamma^{a}_{b c} \dot \phi_0^b X^c $ is a covariant time derivative, with $\Gamma^{a}_{b c} =  \gamma^{a d} (\partial_b \gamma_{d c}  + \partial_c \gamma_{b d} - \partial_d \gamma_{b c} )/2$. The Friedmann equation $3 H^2 = \dot{\phi_0^2}/2 +V$, with $\dot \phi_0^2 \equiv \gamma_{a b} \dot \phi_0^a \dot \phi_0^b$, leads to $\dot H = - \dot \phi_0^2 / 2$. We define orthogonal unit vectors $T^a$ and $N^a$ tangent and normal to the trajectory~\cite{decomposition} as
$T^a = \dot \phi_0^a / \dot \phi_0$ and $N_a = \sqrt{\det\gamma}\epsilon_{a b} T^{b}$, where $\epsilon_{a b}$ is the Levi-Civita symbol with $\epsilon_{12} = 1$. Projecting the background equation of motion along $T^a$ yields $\ddot{\phi}_0+3H\dot{\phi_0}+V_{T}=0$,
where $V_{T} \equiv T^a V_a$. Just as in single-field inflation, we may define the slow-roll parameters $\epsilon \equiv  -\dot{H}/H^2$ and  $\eta_{\parallel} \equiv - \ddot{\phi_0} / \big(H\dot{\phi_0}\big)$. The conditions $\epsilon \ll 1$  and $|\eta_{||}| \ll 1$ ensure that $H$ evolves adiabatically for sufficiently long.  
Projecting along $N^a$, one obtains $D_t T^a = - \dot \theta  N^a$, where $\dot\theta\equiv V_N/\dot\phi_0$ (with $V_N \equiv N^a V_a$) is the angular velocity described by the bends of the trajectory.

We now consider the dynamics of scalar perturbations $\delta\phi^a(t,\mathbi{x}) = \phi^a(t,\mathbi{x}) -\phi_0^a(t)$. We work in the flat gauge and define the comoving curvature and heavy isocurvature perturbations as $\mathcal{R} \equiv -\big(H/\dot \phi\big)  T_a \delta\phi^a$ and $\mathcal{F} \equiv  N_a  \delta\phi^a$, respectively. (A definition of $\mathcal R$ and $\mathcal F$ valid to all orders in perturbation theory is given in \cite{Achucarro:2012sm}). The quadratic order action for these perturbations is 
\begin{equation}\label{total-action-R-F}
S_{\rm 2} = \frac{1}{2} \int \! a^3 \left[  \frac{\dot \phi_0^2}{H^2} \dot {\mathcal R}  ^2  - \frac{\dot \phi_0^2}{H^2} \frac{(\nabla \mathcal R)^2}{a^2}  +   \dot {\mathcal F}  ^2  
- \frac{ (\nabla \mathcal F)^2 }{a^2}  - M_{\rm eff}^2 \mathcal F^2 
-  4  \dot \theta \frac{\dot \phi_0}{H}  \dot{\mathcal R} \mathcal F   \right] \, .
\end{equation}
Here $M_{\rm eff}$ is the effective mass of $\mathcal F$ given by
\begin{equation}\label{meff}
M_{\rm eff}^2 = m^2 - \dot \theta^2 \, ,
\end{equation}
where $m^2 \equiv V_{NN} + \epsilon H^2 \mathbb{R}$, 
with $V_{N N} \equiv N^a N^a \nabla_a \nabla_b V$, and $\mathbb{R}$ is the Ricci 
scalar of the sigma model metric $\gamma_{ab}$.
Notice that $\dot \theta$ couples both fields and reduces the effective mass, suggesting a breakdown of the hierarchy that permits a single field effective description as $\dot \theta^2 \sim  m^2$. As we are about to see, this expectation is somewhat premature. The linear equations of motion in Fourier space are
\begin{align}
\label{eq-of-motion-R}
\ddot {\mathcal R} +  (3  + 2 \epsilon - 2 \eta_{||}) H \dot  {\mathcal R} + \frac{k^2  }{a^2}  {\mathcal R}  = &  2  \dot \theta  \frac{H}{\dot \phi_0} \left[ \dot {\mathcal F} + \bigg( 3 - \eta_{||}  - \epsilon + \frac{\ddot\theta}{H\dot\theta} \bigg) H {\mathcal F} \right] \, ,  
\\
\label{eq-of-motion-F}
\ddot {\mathcal F} +  3  H \dot  {\mathcal F} + \frac{k^2 }{a^2}  {\mathcal F}  + M_{\rm eff}^2  {\mathcal F}  = & - 2 \dot \theta \frac{\dot \phi_0}{H}  \dot {\mathcal R}  \, .
\end{align}
Note that $\mathcal R =$~constant and $\mathcal F = 0$ are non-trivial solutions
 to these equations for arbitrary $\dot \theta$. Since $\mathcal F$ is heavy, $\mathcal F \to 0$ shortly after horizon exit, and $\mathcal R\to$~constant.

We are interested in (\ref{eq-of-motion-R}) and~(\ref{eq-of-motion-F}) in the particular case where $\dot \theta$
is constant and much greater than $M_{\rm eff}$. We first consider the short
wavelength limit where we can disregard  Hubble friction terms and
take $\dot \phi_0 / H$ as a constant. In this regime, the physical
wavenumber $p \equiv k/a$ may be taken to be constant, and
(\ref{eq-of-motion-R}) and (\ref{eq-of-motion-F}) 
simplify to
\begin{equation}\label{eq-of-motion-RF-2}
\begin{split}
\ddot {\mathcal R}_c  + p^2 \mathcal R_c = & + 2 \dot \theta \dot {\mathcal F} \, ,   
\\
\ddot {\mathcal F} + p^2 \mathcal F  + M_{\rm eff}^2  \mathcal F = & - 2 \dot \theta \dot {\mathcal R}_c \, , 
\end{split}
\end{equation}
where we have defined $\mathcal R_c = \big( \dot \phi_0 / H \big) \mathcal R$. The solutions to these equations are found to be~\cite{Achucarro:2010jv}
\begin{equation}\label{solution-RF}
\begin{split}
\mathcal R_c = & \mathcal R_+ e^{i \omega_+ t}  + \mathcal R_- e^{i \omega_- t} \, , 
\\
\mathcal F = &   \mathcal F_+ e^{i \omega_+ t}  + \mathcal F_- e^{i \omega_- t} \, ,  
\end{split}
\end{equation}
where the two frequencies $\omega_{-}$ and $\omega_{+}$ are given by
\begin{equation}\label{frequencies}
 \omega^{2}_{\pm} = \frac{M_{\rm eff}^2 }{2 c_s^2} +  p^2   \pm \frac{M_{\rm eff}^2 }{2  c_s^{2}}  \sqrt{  1 +  \frac{4 p^2 ( 1 - c_s^{2}) }{M_{\rm eff}^2 c_s^{-2}}   } \, , 
\end{equation}
with $c_s$ given by (\ref{speed-of-sound}).
The pairs $( \mathcal R_- , \mathcal F_-)$ and $( \mathcal R_+ , \mathcal F_+)$ represent the amplitudes of both low and high frequency modes respectively, and satisfy
\begin{equation}\label{amplitudes}
\mathcal F_-  =    \frac{ - 2 i \dot \theta \omega_- }{ M_{\rm eff}^2 + p^2  - \omega_-^2 } \mathcal R_-  \, , \quad  \mathcal R_+  =  \frac{-2 i \dot \theta \omega_+}{\omega_+^2 - p^2} \mathcal F_+ \, .
\end{equation}
Thus the fields in each pair oscillate coherently. Of course, we may only neglect the friction terms if both frequencies
satisfy $H \ll \omega_{\pm}$. This implies $H \ll p c_s $, which is what is meant by short wavelength regime.
Integrating out the heavy mode consists in ensuring that the high
frequency degrees of freedom do not participate in the dynamics of the
adiabatic modes. This can only be done in a sensible way if there is a
hierarchy of the form $\omega_{-}^2 \ll \omega_{+}^2$, which given
(\ref{frequencies}) necessarily requires
\begin{equation}\label{range-easy-solution}
p^2  \ll M_{\rm eff }^2 c_s^{-2} \, . 
\end{equation}
This defines the regime of validity of the EFT, in which one has $\omega_{+}^2 \simeq M_{\rm eff}^2  c_s^{-2} = m^2 + 3 \dot\theta^2$ and $\omega_{-}^2 \simeq p^2 c_s^2 + (1 - c_s^2)^2 p^4 / (M_{\rm eff}^2  c_s^{-2})$, and one can clearly distinguish between low and high energy degrees of freedom. Notice that $\omega_{-}^2 \simeq p^2 c_s^2$ for $p^2 \ll M_{\rm eff}^2 / (1 - c_s^2)^2$, and $\omega_{-}^2 \simeq (1 - c_s^2)^2 p^4 / (M_{\rm eff}^2  c_s^{-2})$ for $M_{\rm eff}^2 / (1 - c_s^2)^2 \ll p^2 \ll M_{\rm eff }^2 c_s^{-2}$, which is only possible if $c_s^2 \ll 1$. Then, we see that  condition (\ref{range-easy-solution}) may be rewritten as $\omega_{-}^2 \ll M_{\rm eff}^2 c_s^{-2}$, in light of which the scale $\omega_{+}^2 \simeq M_{\rm eff}^2 c_s^{-2}$ evidently cuts off the low energy regime. One can also re-express (\ref{range-easy-solution}) using (\ref{speed-of-sound}) and (\ref{meff}) as 
$p^2  \ll  4 m^2  / ( 3 c_s^2 + 1 )$. 
From this, we see that contrary to the naive expectation
based on $M_{\rm eff}$, the range of comoving momenta for low energy modes {\it actually increases as the speed of sound decreases}. 
\footnote{
This is consistent with the fact that the heavy eigenvalue of the mass matrix at a particular point along the trough of the potential (in contrast to $M^2_{\rm  eff}$) increases the more the trough deviates from a geodesic of the sigma model metric~\cite{ABHGPP}.} 
  Furthermore, upon quantization~\cite{Achucarro:2010jv} one finds $| \mathcal R_- |^2 \sim c_s^2 / (2
\omega_{-})$ and $|\mathcal F_+ |^2 \sim 1/(2 \omega_{+})$,
implying that high frequency modes are relatively suppressed in
amplitude. Thus, we can safely consider only low frequency modes,
in which case $\mathcal F$ is completely determined by
$\mathcal R_c$ as $\mathcal F = - 2 \dot \theta \dot\calR_c/ \big(
M_{\rm eff}^2 + p^2 - \omega_{-}^2 \big) $. Notice that $\omega_{-}^2 \ll M_{\rm eff}^2 + p^2$, so $ \omega_{-}^2$ may be disregarded here.

As linear perturbations evolve, their physical wavenumber $p \equiv k/a$ decrease and the modes enter the long wavelength regime $ p^2 c_s^2 \lesssim H^2$, where they become strongly influenced by the background and no longer have a simple oscillatory behavior. Now the low energy contributions to $\mathcal F$ satisfy $\dot {\mathcal F} \sim H \mathcal F$, and because $H^2 \ll M_{\rm eff}^2$, we can simply neglect time derivatives in~(\ref{eq-of-motion-F}). On the other hand, high energy modes continue to evolve independently of the low energy modes, diluting rapidly as they redshift. Thus for the entire low energy regime (\ref{range-easy-solution}), time derivatives 
of $\mathcal F$ can be ignored in (\ref{eq-of-motion-F}) and $\mathcal F$ may be solved in terms of $\dot {\mathcal R}$ as
\begin{equation}\label{F-algebraic-R}
{\mathcal F} = -\frac{\dot \phi_0}{H} \frac{2 \dot \theta  \dot {\mathcal R}}{ k^2 / a^2 + M_{\rm eff}^2 } \, . 
\end{equation}
Replacing (\ref{F-algebraic-R}) into (\ref{total-action-R-F}) 
gives the tree level effective action for the curvature perturbation. 
To quadratic order~\cite{Achucarro:2012sm}:
\begin{equation}\label{total-action-R-effective}
S_{\rm eff} =  \frac{1}{2} \int \! a^3 \frac{\dot \phi_0^2}{H^2} \left[ \frac{ \dot {\mathcal R}  ^2 }{c_s^2(k)}  
- \frac{ k^2 \mathcal R^2 }{a^2}   \right] \, ,
\end{equation}
where $c_s^{-2}(k) = 1 + 4 \dot \theta^2 /\big( k^2 / a^2 + M_{\rm eff}^2 \big)$. This $k$-dependent speed of sound is consistent with the modified dispersion relation $\omega_{-}^2 = p^2 c_s^2 + (1 - c_s^2)^2 p^4 / (M_{\rm eff}^2  c_s^{-2})$, where $c_s$ is given by (\ref{speed-of-sound}).
Ref.~\cite{Cespedes:2012hu} studied the validity of (\ref{total-action-R-effective}) in the case where  turns appear suddenly. Consistent with the present analysis, it was found that this EFT is valid even when $\dot \theta^2 \gg M_{\rm eff}^2$, {\it provided the adiabaticity condition} 
\begin{equation}
\bigg| \, \frac{\ddot\theta }{ \dot\theta} \, \bigg| \ll M_{\rm eff}, \label{adiab-cond}
\end{equation}
{\it is satisfied}. This condition states that the turn's angular {\it acceleration} must remain small in comparison to the masses of heavy modes, which otherwise would be  excited. The above straightforwardly implies the more colloquial adiabaticity condition $|\dot \omega_{+} / \omega_{+}^2 | \ll 1$.

We now outline four crucial points that underpin our conclusions: 
\begin{enumerate}
\item The mixing between fields $\mathcal R$ and $\mathcal F$, and modes with frequencies $\omega_{-}$ and $\omega_{+}$ is {\it inevitable} when the background trajectory bends. If one attempts a rotation in field space in order to uniquely associate fields with frequency modes, the rotation matrix would depend on the scale $p$, implying a non-local redefinition of the fields. 

\item Even in the absence of excited high frequency modes, the heavy field $\mathcal F$ is forced to oscillate in pace with the light field $\mathcal R$ at a frequency $\omega_{-}$, so $\mathcal F$ continues to participate in the low energy dynamics of the curvature perturbations. 

\item When $\dot \theta^2 \gg M_{\rm eff}^2$, the high and low energy frequencies become $\omega_{+}^2 \simeq M_{\rm eff}^2  c_s^{-2} \sim 4 \dot \theta^2$ and $\omega_{-}^2 \simeq  p^2 ( M_{\rm eff}^2 + p^2 ) / (4 \dot \theta^2)  $. Thus the gap between low and high energy degrees of freedom is amplified, and one can consistently ignore high energy degrees of freedom in the low energy EFT. 

\item In the low energy regime, the field $\mathcal F$ exchanges kinetic energy with $\mathcal R$ resulting in a reduction in the speed of sound $c_s$ of $\mathcal R$, the magnitude of which depends on the strength of the kinetic coupling $\dot \theta$. This process is adiabatic and consistent with the usual notion of decoupling in the low energy regime (\ref{range-easy-solution}), as implied by (\ref{adiab-cond}).
\end{enumerate}

At the core of these four observations is the simple fact that in time-dependent backgrounds, the eigenmodes and eigenvalues of the mass matrix along the trajectory do not necessarily coincide with the curvature and isocurvature fluctuations and their characteristic frequencies. With this in mind, it is possible to state more clearly the refined sense in which decoupling is operative: {\it  while the fields $\mathcal R$ and $\mathcal F$ inevitably remain coupled, high and low energy degrees of freedom effectively decouple}.

We now briefly address the evolution of modes in the ultraviolet (UV) regime $ p^2 \gtrsim M_{\rm eff }^2  c_s^{-2}$. Here both modes have similar amplitudes and frequencies, and so in principle could interact via relevant couplings beyond linear order (which are proportional to $\dot \theta$). Because these interactions must allow for the non-trivial solutions $\mathcal R =$ constant and $\mathcal F=0$ (a consequence of the background time re-parametrization invariance), their action is very constrained~\cite{Achucarro:2012sm}. Moreover, in the regime $p^2 \gg M_{\rm eff }^2  c_s^{-2}$ the coupling $\dot \theta$ becomes negligible when compared to $p$, and one necessarily recovers a very weakly coupled set of modes, whose  $p\to \infty$ limit completely decouples $\mathcal R$ from $\mathcal F$. This can already be seen in (\ref{F-algebraic-R}),  where contributions to the effective action for the adiabatic mode at large momenta from having integrated out $\mathcal F$, are extremely suppressed for $k^2/a^2 \gg M_{\rm eff }^2$, leading to high frequency contributions to (\ref{total-action-R-effective}) with $c_s = 1$.


We now analyze a model of slow-roll inflation that executes a constant turn in field space, implying an almost constant, suppressed speed of sound for the adiabatic mode. Take fields $\phi^1 = \theta$, $\phi^2 = \rho$ with a metric 
$\gamma_{\theta \theta } = \rho^2$,  $\gamma_{ \rho \rho} = 1$, $\gamma_{ \rho \theta} = \gamma_{\theta \rho} = 0$
(thus
$\Gamma^{\theta}_{ \rho \theta} = \Gamma^{\theta}_{\theta  \rho }  = 1/\rho$ and $\Gamma^{ \rho}_{\theta \theta} = -  \rho $), and potential
\begin{equation}\label{potential}
V(\theta ,  \rho) = V_0 - \alpha \theta + \frac{1}{2}m^2 ( \rho -  \rho_0)^2  \, .
\end{equation}
This model would have a shift symmetry along the $\theta$ direction were it not broken by a non-vanishing $\alpha$. This model is a simplified version of one studied in \cite{Chen:2009we}, where the focus instead was on the regime $M_{\rm eff} \sim m \sim H$ (see also \cite{Chen:2012ge} where the limit $M_{\rm eff}^2 \gg  H^2 \gg \dot \theta^2 $ is analyzed). The background equations of motion are
\begin{equation}
\begin{split}
\ddot \theta + 3 H \dot \theta +  2\dot \theta  \frac{ \dot  \rho }{ \rho }  = & \frac{ \alpha }{ \rho^2} \, , 
\\
\ddot  \rho + 3 H \dot  \rho +   \rho \left( m^2  - \dot \theta^2 \right)  = & m^2   \rho_0 \, . 
\end{split}
\end{equation}
The slow-roll attractor is such that $\dot  \rho$, $\ddot  \rho$ and $\ddot \theta$ are negligible. This means that $H$, $ \rho$ and $\dot \theta$ remain nearly constant and satisfy the following algebraic equations near $\theta = 0$
\begin{equation}\label{algebraic-equations}
\begin{split}
 3 H \dot \theta  = & \frac{\alpha}{\rho^2} \, ,
\\
\dot \theta^2   = &  m^2 \left( 1  -  \frac{ \rho_0 }{ \rho } \right) \, ,   
\\
3 H^2 = & \frac{1 }{2}  \rho^2 \dot \theta^2 + V_0  +  \frac{1}{2} m^2 ( \rho -  \rho_0)^2 \, .
\end{split}
\end{equation}
These equations describe circular motion with a radius of curvature $ \rho$ and angular velocity $\dot \theta$. Here  $M_{\rm eff}^2  = m^2 - \dot \theta^2 $, implying the strict bound $m^2 > \dot \theta^2$. Thus the only way to obtain a suppressed speed of sound is if $\dot \theta^2 \simeq m^2$. 
Our aim is to find the parameter ranges such that the background attractor satisfies $\epsilon \ll 1$, $c_s^2 \ll 1$ and $H^2 \ll M_{\rm eff}^2$ simultaneously. This is given by
\begin{equation}\label{inequalities}
1 \gg \frac{\rho_0}{4} \left( \frac{m \sqrt{3 V_0}}{ \alpha} \right)^{1/2} \gg \frac{V_0}{6 m^2} \gg  \frac{ \alpha} { 4 \sqrt{3 V_0} m } \  . 
\end{equation}
If these hierarchies are satisfied, the solutions to (\ref{algebraic-equations}) are well approximated by
\begin{equation}\label{aprox-background-sol}
\rho^2= \frac{\alpha}{\sqrt{3 V_0}  m} \, , \quad \dot \theta = m - \frac{m  \rho_0}{2}\left( \frac{m \sqrt{3 V_0}}{ \alpha}  \right)^{1/2} \, , 
\end{equation}
and $ H^2 = V_0/3$,  up to fractional corrections of order $\epsilon$, $c_s^2$ and $H^2 / M_{\rm eff}^2$. We note that the first inequality in (\ref{inequalities}) implies $ \rho \gg  \rho_0$, and so the trajectory is displaced off the adiabatic minimum at $ \rho_0$. However, the contribution to the total potential energy implied by this displacement is negligible compared to $V_0$. After $n$ cycles around $ \rho=0$ one has $\Delta \theta =  2 \pi n$, and the value of $V_0$ has to be adjusted to $V_0  \to V_0 - 2 \pi n \alpha$. This modifies the expressions in (\ref{aprox-background-sol}) accordingly, and allows us to easily compute the adiabatic variation of certain quantities, such as
$s \equiv \dot c_s / (c_s H) =  - \epsilon / 4$, and $\eta_{||} = - \epsilon / 2$, where $\epsilon = \sqrt{3} \alpha m^2 / (2 V_0^{3/2})$.
These values imply a spectral index $n_{\mathcal R}$ for the power spectrum $\mathcal P_{\mathcal R} = H^2 / (8 \pi^2 \epsilon c_s)$ given by $n_{\mathcal R} -1=  -4\epsilon + 2 \eta_{||} - s =  - 19 \epsilon / 4 $.
It is now possible to find reasonable values of the parameters in such a way that observational bounds are satisfied. Using (\ref{aprox-background-sol}) we can relate the values of $V_0$, $\alpha$, $m$ and $ \rho_0$ to the measured values $\mathcal P_{\mathcal R} $ and $n_{\mathcal R}$, and to hypothetical values for $c_s$ and $\beta \equiv H / M_{\rm eff}$ as
\begin{equation}\label{par}
\begin{split} 
V_0 =& \frac{96}{19}  \pi^2 (1 - n_{\mathcal R})  \mathcal P_{\mathcal R}c_s \, , 
\\
m^2 =&  \frac{8}{19}  \pi^2 (1 - n_{\mathcal R})  \mathcal P_{\mathcal R} c_s^{-1} \beta^{-2} \, ,
\\
\alpha =&  6 \left( \frac{16}{19} \right)^2 \pi^2  (1 - n_{\mathcal R})^2  \mathcal P_{\mathcal R} c_s^2  \beta \, , 
\\
\rho_0   =&  16 c_s^{3}   \beta \sqrt{ \frac{2}{19} (1-n_{\mathcal R}) } \,    .  
\end{split}
\end{equation}
Following WMAP7, we take $\mathcal P_{\mathcal R}  = 2.42  \times 10^{-9}$ and  $n_{\mathcal R} = 0.98$~\cite{Komatsu:2010fb}. Then, as an application of relations (\ref{par}), we look for parameters such that
\begin{equation}\label{desired-values}
  c_s^2 \simeq 0.06 \, , \quad M_{\rm eff}^2  \simeq 250 H^2 \, ,  
\end{equation}
(which imply $H^2 \simeq  1.4 \times 10^{-10}$), according to which $V_0 \simeq 5.9 \times 10^{-10}$, $\alpha \simeq 1.5 \times 10^{-13}$, $m \simeq 4.5 \times 10^{-4}$ and $ \rho_0 \simeq 6.8 \times 10^{-3}$,
from which we note that $m$, $\rho_0$ and $\alpha^{1/4}$ are naturally all of the same order. We have checked numerically that the background equations of motion are indeed well approximated by (\ref{aprox-background-sol}), up to fractional corrections of order $c_s^2$. More importantly, we obtain the same nearly scale invariant power spectrum $ \mathcal P_{\mathcal R} $ using both the full two-field theory described by (\ref{eq-of-motion-R}) and~(\ref{eq-of-motion-F}), and the single field EFT described by the action~(\ref{total-action-R-effective}). The evolution of curvature perturbations in the EFT compared to the full two-field theory for the long wavelength modes is almost indistinguishable given the effectiveness with which (\ref{range-easy-solution}) is satisfied, with a marginal difference $\Delta \calP_\calR / \calP_\calR \simeq 0.008$. This is of order $ (1 - c_s^2) H^2 / M_{\rm eff}^2$, which is consistent with the analysis of Ref.~\cite{Cespedes:2012hu}. Despite the suppressed speed of sound in this model, a fairly large tensor-to-scalar ratio of $r = 16 \epsilon c_s \simeq 0.020$ is predicted.

As expected, for $c_s^2\ll 1$ a sizable value of $f_{\rm NL}^{\rm (eq)}$ is implied. The cubic interactions leading to this were deduced in Ref.~\cite{Achucarro:2012sm} which for constant turns is given by~\cite{bispectrum}
\begin{equation}
f_{\rm NL}^{\rm (eq)} = \frac{125}{108}\frac{\epsilon}{c_s^2} + \frac{5}{81}\frac{c_s^2}{2}\left( 1-\frac{1}{c_s^2} \right)^2 + \frac{35}{108}\left( 1-\frac{1}{c_s^2} \right)  .
\end{equation}
This result is valid for any single-field system with constant $c_s$ obtained by having integrated out a heavy field. Recalling that the spectral index $n_T$ of tensor modes is $n_T = - 2\epsilon$, for $c_s\ll 1$ we find a consistency relation between three potentially observable parameters, given by $\fnl^{\rm (eq)} = - 20.74 \, n_T^2 / r^2 $.
In the specific case of the values in (\ref{desired-values}), we have $f_{\rm NL}^{\rm (eq)} \simeq -4.0$. This value is both large and negative, so future observations could constrain this type of scenario. Finally, one can ask if the EFT corresponding to (\ref{desired-values}) remains weakly coupled throughout. For this, one needs to satisfy~\cite{Achucarro:2012sm} $ \omega_{-}^4 < \Lambda_{\rm sc}^4$, where $\Lambda_{\rm sc}^4 \simeq 4\pi \epsilon H^2 c_s^{5} / (1-c_s^2)$ is the strong coupling scale~\cite{Cheung:2007st,Baumann:2011su}. In order to assess this, we notice that for $c_s^2 \ll 1$, values of $\omega_{-}^2$ of order $M_{\rm eff}^2$ are well within the low energy regime ($M_{\rm eff}^2 \ll \omega_{+}^2$) and denotes a reference energy scale up to which we may trust the low energy EFT. We find $M_{\rm eff}^4 / \Lambda_{\rm sc}^4 \simeq 0.18$. Furthermore, although we did not address how inflation ends, the choice (\ref{desired-values}) allows for at least $45$ $e$-folds of inflation, necessary to solve the horizon and flatness problems. We stress that various other values can be chosen in (\ref{desired-values}) to arrive at similar conclusions. For example, requiring $35$ $e$-folds with $M^2_{\rm eff} \simeq 100 H^2$, $c_s^2 \simeq 0.02$, implies $V_0 \simeq 3.4 \times 10^{-10}$, $\alpha \simeq 8.1 \times 10^{-13}$, 
$m \simeq 3.8 \times 10^{-4}$, $\rho_0 \simeq 2.1 \times 10^{-4}$,  so that the strong coupling scale becomes $M^4_{\rm eff}/\Lambda^4_{\rm sc} \simeq 0.34$. In this case we find
$f_{\rm NL}^{\rm (equil)}  \simeq -14 .$
 As an illustrative limit, we can try to saturate the strong coupling bound given a particular hierarchy between $H$ and $M_{\rm eff}$. In doing so, we entertain the situation where our model approximates the dynamics of inflation only over the range where modes accessible to us by observations had exited the horizon. Requiring $M_{\rm eff}^2 / H^2  \simeq 100$ and $\Lambda_{\rm sc} \simeq M_{\rm eff}$, we find that approximately $14$ $e$-folds can be generated where the speed of sound is reduced to $c_s^2 \simeq 0.01$.

In summary, the active ingredients of this toy example are rather
minimal and may well parametrize a generic class of inflationary
models, such as axion-driven inflationary scenarios in string theory. Our results complement those of
Ref.~\cite{Tolley:2009fg,Achucarro:2010jv,Achucarro:2010da,Cespedes:2012hu,Achucarro:2012sm}
and emphasize the refined sense in which EFT techniques are applicable
during slow-roll inflation~\cite{Cheung:2007st,infEFT}. In particular,
contrary to the standard perspective regarding the role of
UV physics during inflation, heavy fields may influence the evolution
of curvature perturbations $\mathcal R$ in a way consistent with
decoupling between low and high energy degrees of freedom.

\subsection*{Acknowledgements}

We thank T. Battefeld, C. Burgess, D. Green, M. Horbatsch and P. Ortiz for useful discussions. 
This work was supported by funds from the Netherlands Foundation for Fundamental Research on Matter (F.O.M), Basque Government grant IT559-10, the Spanish Ministry of Science and Technology grant FPA2009-10612 and Consolider-ingenio programme CDS2007-00042 (AA), by a  Leiden Huygens Fellowship (VA), Conicyt under the Fondecyt initiation on research  project 11090279  (SC \& GAP), a Korean-CERN fellowship (JG) and the CEFIPRA/IFCPAR project 4104-2 and ERC Advanced Investigator Grants no. 226371 ``Mass Hierarchy and Particle Physics at the TeV Scale'' (MassTeV) (SP).
We thank King's College London, University of Cambridge (DAMTP), CPHT at Ecole Polytechnique and Universiteit Leiden for their hospitality.

\end{document}